\documentclass[pdflatex,sn-nature]{sn-jnl}

\usepackage{graphicx}%
\usepackage{multirow}%
\usepackage{amsmath,amssymb,amsfonts}%
\usepackage{amsthm}%
\usepackage{mathrsfs}%
\usepackage[title]{appendix}%
\usepackage{xcolor}%
\usepackage{textcomp}%
\usepackage{manyfoot}%
\usepackage{booktabs}%
\usepackage{algorithm}%
\usepackage{algorithmicx}%
\usepackage{algpseudocode}%
\usepackage{listings}%

\theoremstyle{thmstyleone}%
\theoremstyle{thmstyletwo}%

\theoremstyle{thmstylethree}%

\begin{document}

\title[Article Title]{Automated design of soft--rigid hybrid robots for dynamic locomotion}

\author*[1]{\fnm{Hiroki} \sur{Kobayashi}}\email{hiroki.kobayashi@mosk.tytlabs.co.jp}

\author[2]{\fnm{Yuki} \sur{Takaha}}
\equalcont{Present address: \orgdiv{Institute of Industrial Science}, \orgname{The University of Tokyo}, \orgaddress{\street{4-6-1 Komaba}, \city{Meguro-ku}, \postcode{153-8505}, \state{Tokyo}, \country{Japan}}}
\author[1]{\fnm{Changyoung} \sur{Yuhn}}
\author[1]{\fnm{Yuki} \sur{Sato}}
\author[1]{\fnm{Sunao} \sur{Tomita}}
\author[1]{\fnm{Atsushi} \sur{Kawamoto}}
\author[1]{\fnm{Tsuyoshi} \sur{Nomura}}

\affil[1]{\orgname{Toyota Central R\&D Labs., Inc.}, \orgaddress{\street{1-4-14 Koraku}, \city{Bunkyo-ku}, \postcode{112-0004}, \state{Tokyo}, \country{Japan}}}

\affil[2]{\orgdiv{Graduate School of Arts and Sciences}, \orgname{The University of Tokyo}, \orgaddress{\street{3-8-1 Komaba}, \city{Meguro-ku}, \postcode{153-8902}, \state{Tokyo}, \country{Japan}}}


\abstract{
Rigid-bodied robots often lack compliance needed to adapt to unstructured environments, while fully soft robots, though highly adaptable, struggle with scalability and load capacity.
In nature, musculoskeletal systems balance strength and flexibility by integrating hard and soft tissues.
Inspired by this principle, we present an automated design method for soft--rigid hybrids that optimizes a freeform soft-body shape, a stiff truss layout, and multi-channel actuation. Our differentiable simulator couples the material point method (MPM) for deformable bodies with extended position-based dynamics (XPBD) for truss elements, enabling gradient-based search. The optimization generates truss skeletons that transmit actuation forces to the soft body.
We fabricate the optimized design and evaluate it on a walking task.
Experiments reproduce the walking mode predicted by the optimization, which does not emerge without the skeleton. Modal analysis further suggests that the skeleton enables deformation modes near the actuation frequency that promote effective stride generation.}

\maketitle

\section*{Introduction}\label{sec1}
Traditional robots have predominantly been built from rigid components such as metal links and motors. 
Such rigid mechanisms have offered great advantages in terms of precision and repeatability, enabling automation across a wide range of industrial tasks~\cite{siciliano2008springer}. 
However, their mechanical rigidity often limits adaptability to uncertain or changing environments, and their operation in close proximity to humans often raises safety and interaction concerns~\cite{rus2015design}. 

In contrast, soft robots have attracted increasing attention as a new paradigm to achieve adaptive and safe interaction with the environment~\cite{kim2013soft,rus2015design}. Constructed from highly deformable and compliant materials, they can passively absorb unexpected contact and conform to complex or uncertain surroundings. Their continuous body deformation allows rich behaviors by relatively simple actuation or control inputs~\cite{kim2013soft}. 
Despite these advantages, the inherent softness of such robots imposes several limitations on their mechanical performance~\cite{fan2024overview}. Their compliant structures cannot generate or transmit large forces efficiently, making it difficult to perform high-load or high-precision tasks.

In the biological world, organisms exhibit diverse structural strategies for locomotion and load bearing, ranging from fully soft-bodied organisms such as worms and jellyfish to musculoskeletal systems that integrate soft tissues with rigid skeletons~\cite{kier1985tongues,alexander2003principles}. 
While biological structures at small scales can rely on compliant materials and distributed deformation,
increasing body size imposes progressively greater mechanical stress due to the scaling of mass with volume~\cite{alexander2003principles,mcmahon1973size}.
These observations suggest that combining compliant tissues with rigid load-bearing elements provides a scalable strategy for balancing flexibility and structural integrity.

The promise of soft–rigid hybrid robotics lies in combining the compliance and adaptability of soft materials with the strength and load-bearing capability of rigid structures. Several hybrid designs have been developed in the context of robotic manipulators~\cite{zhu2022soft,bern2022simulation,patterson2025design,xia2025topology}. For example, soft robotic fingers with rigid bones and tendon-driven actuation embedded in compliant bodies have been shown to improve load transmission compared with purely soft structures, while still retaining large deformation capability~\cite{bern2022simulation}. Beyond manipulation tasks, hybrid structures have also been explored in legged robots, where rigid segments are connected by compliant backbones to simplify control while maintaining structural integrity~\cite{yaman2026maneuverable}. In addition, lattice-based hybrid structures have been proposed to locally modulate stiffness, enabling both locomotion and manipulation capabilities in elephant-inspired robots~\cite{guan2025lattice}.

Several works have discussed design principles and strategies for constructing soft–rigid hybrid manipulators~\cite{culha2016design,fan2024overview}.
These studies emphasize that the performance of hybrid systems depends critically on how rigid load-bearing elements and compliant regions are arranged, as they play distinct mechanical roles.
Soft materials enable large, continuous deformation and can realize rich behaviors through distributed deformation modes, but their response is inherently high-dimensional and strongly dependent on geometry~\cite{rus2015design}.
In contrast, skeletal structures can efficiently transmit forces over distance and provide structural support with relatively small mass, making the morphology of rigid elements a key factor in determining overall system behavior~\cite{smith2015effect,chen2019improving}.

Despite these insights, determining effective configurations of soft and rigid components remains challenging.
In practice, the design of soft–rigid hybrid systems is often guided by biological inspiration~\cite{zhu2022soft,bern2022simulation,guan2025lattice}, which provides valuable insights into effective morphological design.
However, translating such principles into concrete designs is not straightforward, and skeletal layouts and body geometries are typically specified manually and refined through iterative processes.
This challenge becomes particularly pronounced in locomotion tasks~\cite{calisti2017fundamentals}, as behavior emerges from interactions between the time-varying body state and the environment.
Because these interactions are strongly coupled with the system dynamics, the resulting motion depends sensitively on the interplay between morphology and dynamics~\cite{pfeifer2006body}, making it difficult to anticipate effective designs without systematic exploration.

Computational approaches have emerged as a promising approach for the design of robotic morphologies and control strategies in complex settings. Early approaches combined rigid-body simulation and evolutionary algorithms to discover locomotion strategies through iterative exploration of body structures and control policies~\cite{sims1994evolving}. This line of research has been extended to more structured design spaces, such as voxel-based representations~\cite{cheney2014unshackling,bhatia2021evolution} and variable geometry truss~\cite{gu2025optimization}. 
Recent advances in computational methods have broadened the range of design approaches, with emerging technologies leveraging machine learning~\cite{diaz2023machine,sun2024machine,qiu2026robomorph} and differentiable physics~\cite{hu_difftaichi_2020,du2021diffpd,stuyck2023diffxpbd} to enable gradient-based optimization in physics-involved design problems.

In soft robotics, differentiable simulation has been applied to the design of soft body shape~\cite{sato2023topology,yuhn20234d,cochevelou2023differentiable,matthews2023efficient,kobayashi2024computational,sato2025computational,wang2025co,hashiguchi_unified_2025}, actuator layout~\cite{yuhn20234d,cochevelou2023differentiable,matthews2023efficient,wang2025co}, and actuation control, including phase modulation~\cite{hu2018moving,du2021diffpd,cochevelou2023differentiable}, frequency modulation~\cite{wang2025co}, and feedforward~\cite{yuhn20234d} or feedback controllers~\cite{sato2025computational}, particularly in the context of locomotion tasks. Several of these designs have been physically fabricated and experimentally validated~\cite{matthews2023efficient,kobayashi2024computational}. Extensions toward soft--rigid design have also begun to emerge in differentiable simulation, for example through approaches that optimize the spatial distribution of soft, stiff, actuated, and void regions~\cite{cochevelou2023differentiable}. However, such approaches still face limitations in the explicit design of soft--rigid hybrid structures with skeletal or mechanism-like components.

These limitations arise from both numerical and representational aspects. 
From a numerical perspective, under explicit time integration schemes, the range of material stiffness is often restricted by time-step constraints, making it difficult to realize the strong stiffness contrast expected between compliant bodies and effectively rigid structural elements. 
From a structural representation standpoint, continuous design representations pose challenges in depicting discrete structural features.
Many existing approaches—either explicitly or implicitly—adopt concepts from topology optimization~\cite{bendsoe1988generating}, where structures are represented as continuous scalar fields. While this representation is well suited for describing deformable bodies, it is less natural for expressing sparse load paths, discrete connections, or mechanism-like force transmission. Although emerging studies have begun to explore such behaviors within continuum frameworks~\cite{sayo2024topology}, these approaches remain limited in capturing the structural characteristics of articulated or skeletal systems, particularly in dynamic settings.

Here, we present a computational framework for the automated design of soft–rigid hybrid robots. Our method represents a robot as a continuous soft body reinforced by a network of stiff truss elements, and simultaneously optimizes both together with the actuation signals to exploit their coupled dynamics. Efficient gradient-based optimization is enabled by a differentiable simulator that couples the material point method (MPM)~\cite{sulsky1994particle}, which has been widely adopted for soft robotic design with differentiable simulation~\cite{sato2023topology,yuhn20234d,cochevelou2023differentiable,matthews2023efficient,kobayashi2024computational,sato2025computational,hashiguchi_unified_2025}, with extended position-based dynamics (XPBD)~\cite{macklin2016xpbd} for truss structures. XPBD extends position-based dynamics (PBD)~\cite{muller2007position}, a method originally developed in computer graphics, by incorporating stiffness control while avoiding the need to solve large global systems, and has recently been extended to a differentiable formulation~\cite{stuyck2023diffxpbd}. 
From a structural representation perspective, discrete representations such as trusses or rigid blocks have been employed in optimization frameworks for linkage mechanisms~\cite{stolpe2005design,kim2014topology,pan2019globally}, making them well suited for problems where connectivity and force transmission are essential. The combination of a soft continuum with an overlaid skeletal truss also relates to layered structural concepts explored in topology optimization, where multiple material or structural layers are coupled to achieve functionalities that cannot be realized by a single component alone~\cite{xia2025topology,zhao2025extreme}.

We apply this framework to a walking task, optimizing soft–rigid morphology for locomotion performance under realistic actuation. The optimized designs are then fabricated and tested in walking experiments to validate the effectiveness of the optimized hybrid morphologies. Beyond performance improvement, the resulting morphologies provide insight into how the roles of soft and rigid components should be distributed across body scales to achieve both flexibility and force transmission.

\begin{figure}[!tbp]
    \centering
    \includegraphics[width=1.0\linewidth]{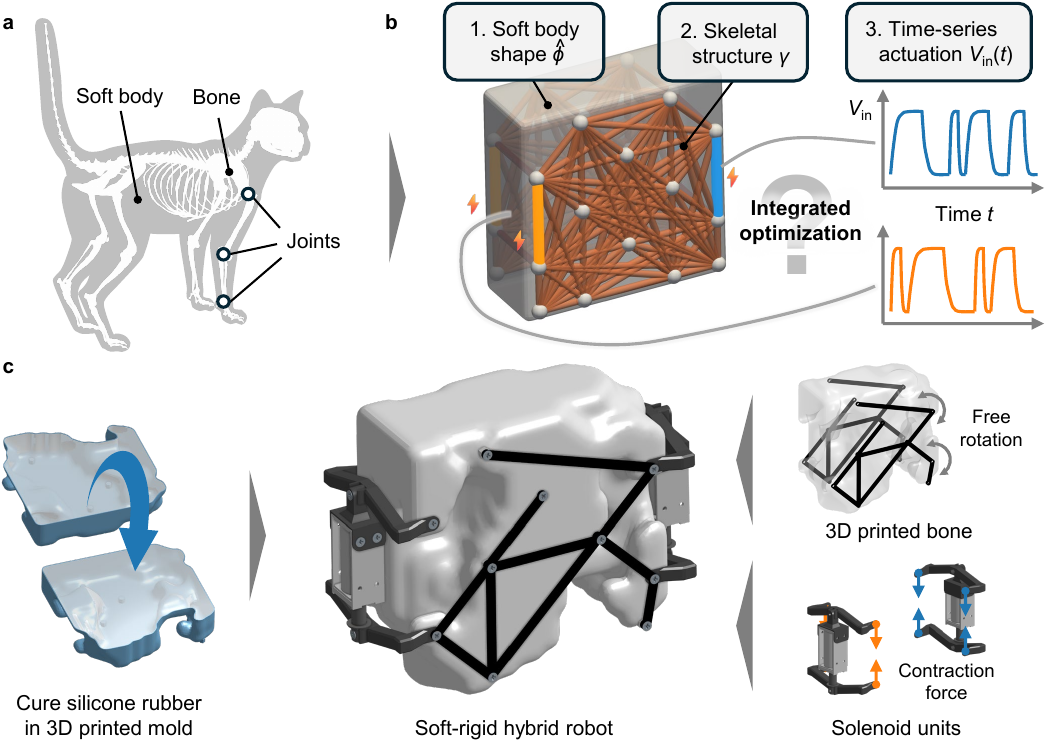}
    \caption{Concept and framework for the design of soft--rigid hybrid robots. \textbf{a} Musculoskeletal system combines compliance for adapting to the environment with strength for supporting body weight and transmitting large forces. \textbf{b} Overview of the proposed design framework, where the soft body morphology, skeletal layout, and time-dependent actuation are co-optimized. The soft body is designed within a rectangular domain, while a set of staggered nodes is placed on both sides of the body to define candidate skeletal connections. The actuation is defined by contraction forces whose timing is optimized. \textbf{c} Fabrication process of soft--rigid hybrid robot. The soft body is molded from silicone rubber, the skeleton is fabricated via 3D printing, and the system is assembled together with solenoids using simple bolted connections.}
    \label{fig1}
\end{figure}

\section*{Results}

\subsection*{Framework for automated design}
Inspired by musculoskeletal systems in animals (Fig.~\ref{fig1}a), we formulate the automated design of soft--rigid hybrid robots as the optimization of the soft body shape, the skeletal truss structure, and the time-series actuation (Fig.~\ref{fig1}b). 
These three components are optimized simultaneously such that the body structure and actuation pattern are coordinated to achieve favorable deformation modes.

The robots are driven by solenoid actuators consisting of a coil and a movable core that can slide freely until contact. The actuation forces are transmitted through the skeletal structure and the soft body.
The objective is defined based on the horizontal locomotion distance of the robot. 
The three types of design variables are updated simultaneously, in an iterative manner, using a gradient-based optimization method to search for an optimal design.

For fabrication, the soft body is molded from silicone rubber, while the skeletal components are 3D printed from polylactic acid (PLA), a commonly used thermoplastic filament.
By attaching the solenoid actuators and the skeletal structure to the soft body using bolted connections, the soft--rigid hybrid robot is assembled (Fig.~\ref{fig1}c). 
This setup allows the optimized body shape, skeleton, and actuation pattern to be directly translated into a physically realizable robot with minimal design modification.

\subsection*{Differentiable simulation of soft--rigid coupled systems}
We modeled the robot as a coupled system consisting of a soft continuum simulated by the material point method (MPM) and an stiff bar network (skeleton and solenoids) simulated by extended position-based dynamics (XPBD). 
In the MPM framework, the soft body is represented as a collection of particles carrying mass, momentum, and deformation states. 
At each time step, these physical quantities are transferred to an Eulerian background grid, where momentum updates and boundary interactions are computed, after which the updated states are transferred back to the particles.
Within this MPM update, the motion of the XPBD nodes is first predicted under actuator forces, and the constraints of the bar network are subsequently resolved.
The resulting momentum is exchanged through the same background grid used by the soft continuum. 
This shared grid representation enables two-way coupling between the deformable soft body and the stiff bar network. 

Both MPM and XPBD are formulated using local operations that avoid global matrix solves, making the simulation highly parallelizable. 
This local update structure also enables efficient implementation with automatic differentiation. 
Contact interactions are handled on the MPM grid, such that the embedded skeleton interacts with the environment through the surrounding soft continuum rather than through explicit rigid-body contact solvers, which are difficult to derive the exact gradient.
For details of the algorithm and discretization, see ``Simulation setup'' in the Methods section.

The soft continuum is simulated using the moving least squares material point method (MLS-MPM,~\cite{hu2018moving}).
A compressible neo-Hookean constitutive law with an additional viscous term~\cite{sato2025computational} was employed to represent dissipative effects. 
For formulations, see "Soft body modeling" in the Methods section. 

In the soft-body topology optimization, the design variable $\phi \in [-1,1]$ was first smoothed using a particle-based filter and then projected through a smooth Heaviside-type function to obtain the density field $\hat{\phi} \in [0,1]$ defining the soft-body geometry, following prior studies~\cite{yuhn20234d,kobayashi2024computational}.
The Lam\'e parameters and mass density were interpolated from the projected density field $\hat{\phi}$ using a SIMP-based formulation~\cite{bendsoe2013topology} with cubic penalization.

Next, the stiff skeleton was represented as a network of bar constraints within the XPBD framework, modeled as a system of mass points (nodes) connected by bars. 
XPBD introduces Lagrange multipliers that enable explicit control of the bar stiffness through a compliance parameter~\cite{macklin2016xpbd}. For formulations, see ``XPBD formulation'' in the Methods section.
In the present formulation, the axial stiffness $\kappa$ was defined per unit length, such that shorter bars exhibit higher effective stiffness. 
The mass $m$ at the nodes was computed by distributing the mass of each bar element equally to its endpoints and summing the contributions from all connected bars. 

For skeletal topology optimization, a design variable $\gamma \in [0,1]$ was assigned to each bar element. 
This variable controlled the axial stiffness $\kappa$ of each bar element, while the corresponding bar mass was distributed equally to its two endpoint nodes. 
The nodal masses were then obtained by summing the contributions from all connected bars.
The bar stiffness and mass contributions were interpolated from the design variable $\gamma$ using a special interpolation function,
\begin{equation}
(m,\, \kappa) \propto
\frac{(\gamma+\epsilon_b)^p-\epsilon_b^p}{(1+\epsilon_b)^p-\epsilon_b^p},
\end{equation}
which was constructed such that the gradient does not vanish at $\gamma=0$, preventing skeletal elements from becoming permanently inactive during optimization. 
Here, $\epsilon_b$ is a small positive parameter ($0.1$), and $p$ is the penalization exponent. 
A larger exponent ($p=6$) than that used for the soft body (cubic) was employed to suppress the effective stiffness of intermediate densities. 
This reduces the stiffness contrast between the soft body and low-density skeletal regions during optimization, while also promoting clearer separation toward either $\gamma = 0$ or $\gamma = 1$ around the threshold $\gamma = 0.5$, thereby preventing structurally important elements from being eliminated. 
In addition, lower bounds of $10^{-9}$ relative to the maximum values were imposed on both mass and stiffness to avoid numerical instability in the XPBD solver.
Because the skeletal bars were slender (Fig.~\ref{fig1}c), a simplified buckling-aware stiffness model was additionally introduced. For detailed formulations, see ``Buckling-aware stiffness model'' in the Methods section.

In addition to the skeletal truss, each solenoid unit was represented within the same XPBD framework using bar elements. 
However, unlike the passive skeletal bars, the actuator bars were modeled to reflect the structural characteristics of the solenoid mechanism, including the sliding motion between the coil assembly and the iron core, as well as the bracket-based attachment structure integrated into the robot body.

Each actuator unit was represented by four bars (Fig.~\ref{fig1}b): two axial bars aligned with the actuation direction and two lateral support bars corresponding to the physical bracket structure shown in Fig.~\ref{fig1}c. 
The axial bars were modeled using a state-dependent stiffness response. 
During free sliding, the bars were assigned very low stiffness, allowing nearly unconstrained extension. 
Once the contraction limit was reached, the stiffness increased substantially to represent contact between the internal actuator components. Detailed formulations of the stiffness switching are provided in the ``Solenoid actuator model''  of the Methods section.

The actuation was represented as external nodal forces rather than prescribed kinematic contraction. 
The generated force depended on both the input voltage and the current actuator length,
\begin{equation}
F_{\mathrm{sol}}(t,\ell)=F_{\max} V_{\mathrm{in}}(t)\,\eta_{\mathrm{stroke}}(\ell),
\label{eqn_stroke_force}
\end{equation}
where $V_{\mathrm{in}}(t)$ denotes the normalized voltage input. 
The voltage input $V_{\mathrm{in}}(t)$ was parameterized using design variables $w_{a,k}\in[0,1]$, representing a pulse sequence in the time domain, where $a$ denotes the actuator index and $k$ represents the pulse index in the temporal actuation sequence. 
The factor $\eta_{\mathrm{stroke}}(\ell)$ represents a stroke-dependent attenuation associated with the engagement of the solenoid core. 
This formulation reproduces the finite stroke and force attenuation characteristics of the physical solenoid actuator. 
Details are provided in the ``Solenoid actuator model''  of the Methods section.

\subsection*{Optimization problem for locomotion task}
We formulate an optimization problem to co-design the soft-body shape, the skeletal structure, and the actuation timing.
The design variables are:
(i) the soft-body variable $\phi_i\in[-1,1]$ defined at each soft particle,
(ii) the skeletal variable $\gamma_s\in[0,1]$ defined for each bone bar, and
(iii) the actuation variables $w_{a,k}\in[0,1]$ that parameterize the input voltage of actuator $a$ over time.
The filtering and projection of $\phi$ follow prior work~\cite{yuhn20234d,kobayashi2024computational} and produce the physical density field $\hat{\phi}\in[0,1]$.
The skeletal density variable $\gamma_s$ is directly mapped to the stiffness of each bar without filtering or projection.
The actuation variables $w_{a,k}$ represent a pulse sequence discretized in timed domain. Based on $w_{a,k}$, a smooth voltage signal $V_{\mathrm{in}}(t)$ is generated via a superposition of Gaussian functions~\cite{yuhn20234d}. 

The primary objective is to maximize the forward locomotion distance of the robot after the initial settling phase under gravity.  
Let $\boldsymbol{x}_{\mathrm{cg}}(t)$ denote the center-of-mass position at time $t$, and $\boldsymbol{e}_{x}$ the unit vector along the prescribed locomotion direction.  
The forward displacement is defined as
\begin{equation}
L_x
=
\big(
\boldsymbol{x}_{\mathrm{cg}}(t_{\mathrm{end}})
-
\boldsymbol{x}_{\mathrm{cg}}(t_{\mathrm{start}})
\big)
\cdot
\boldsymbol{e}_x,
\end{equation}
where $t_{\mathrm{start}}$ and $t_{\mathrm{end}}$ denote the start and end times of actuation, respectively.  
The interval before $t_{\mathrm{start}}$ is introduced to allow the structure to settle into a stable configuration under gravity and contact.

To suppress undesirable deformation modes such as collapse under gravity, falling during locomotion, and lateral or vertical drifting, we penalize deviations from rigid-body translation along the locomotion direction.  
Unlike the locomotion objective, the penalty is measured from the initial configuration at $t=0$ to suppress deformation caused by self-weight before actuation begins.
For each particle or skeletal node $i$, the residual displacement is defined as
\begin{equation}
\tilde{\boldsymbol d}_i
=
\big(
\boldsymbol{x}_i(t_{\mathrm{end}})
-
\boldsymbol{x}_i(0)
\big)
-
\big(
x_{\mathrm{cg},x}(t_{\mathrm{end}})
-
x_{\mathrm{cg},x}(0)
\big)\boldsymbol e_x,
\label{eqn_deviation}
\end{equation}
where $x_{\mathrm{cg},x}$ denotes the $x$ coordinate of the center of gravity.  

This residual measures the deviation from ideal rigid translation in the $x$ direction, and therefore penalizes structural deformation as well as motion components in the $y$ and $z$ directions.  
The non-uniform motion measures, $D_{\mathrm{soft}}$ and $D_{\mathrm{bone}}$, are defined as the mass-weighted averages of the Euclidean norms $\|\tilde{\boldsymbol d}_i\|_2$ over the corresponding particles and skeletal nodes.

To enhance near-binary designs, we penalize intermediate values of all design variables using a unified quadratic relaxation. Specifically, for the soft material densities $\hat{\phi}_i$, skeletal densities $\gamma_s$, and actuation variables $w_{a,k}$, we define the constraint functions $C_{\mathrm{soft}}$, $C_{\mathrm{bone}}$, and $C_{\mathrm{act}}$ as the corresponding averaged quantities $\langle \hat{\phi}_i(1-\hat{\phi}_i) \rangle$, $\langle \gamma_s(1-\gamma_s) \rangle$, and $\langle w_{a,k}(1-w_{a,k}) \rangle$ minus prescribed tolerances.

In addition, we define $C_{N_{\mathrm{bone}}}$ based on the average skeletal density $\langle \gamma_s \rangle$ to control the number of bone elements, where the averaging is taken over designable bone bars excluding actuator and bridge elements.

All objective and constraint terms are evaluated through the differentiable MPM--XPBD simulation, and the design variables are updated simultaneously using Adam optimizer~\cite{kingma_adam_2015} with an augmented Lagrangian formulation. 
The overall objective is defined as the augmented Lagrangian,
\begin{equation}
\mathcal{L}
=
- L_{\mathrm{x}}\,
\frac{\bar{D}_{\mathrm{soft}}}{\bar{D}_{\mathrm{soft}}+D_{\mathrm{soft}}}\,
\frac{\bar{D}_{\mathrm{bone}}}{\bar{D}_{\mathrm{bone}}+D_{\mathrm{bone}}}
\;+\;
\sum_{i}
\left(
- \lambda_i C_i
+ \tfrac{1}{2}\sigma_i C_i^2
\right),
\end{equation}
where the first term represents the primary objective $L_{\mathrm{x}}$, combined with penalties on the non-uniform motion of the soft body $D_{\mathrm{soft}}$ and the skeleton $D_{\mathrm{bone}}$. 
The reference parameters $\bar{D}_{\mathrm{soft}}$ and $\bar{D}_{\mathrm{bone}}$ determine the characteristic scales of these penalties and are set to $0.005$ in the present study. 
When $D=\bar{D}$, the corresponding factor becomes $0.5$, thereby reducing the locomotion objective by half. 
The constraints $C_i \in \{C_{\mathrm{soft}}, C_{\mathrm{bone}}, C_{\mathrm{act}}, C_{N_{\mathrm{bone}}}\}$ are defined such that $C_i>0$ indicates violation.
Here, the constraint $C_{\mathrm{soft}}$ is activated only after the skeletal constraint $C_{\mathrm{bone}}$ has been satisfied.
This choice is motivated by the different roles of the two components: changes in the skeletal structure have a dominant impact on the overall mechanism, and enforcing $C_{\mathrm{soft}}$ too early can overly restrict the design flexibility of the soft body.
\subsection*{Automated design process and behaviors}
\begin{figure}[!tbp]
    \centering
    \includegraphics[width=1.0\linewidth]{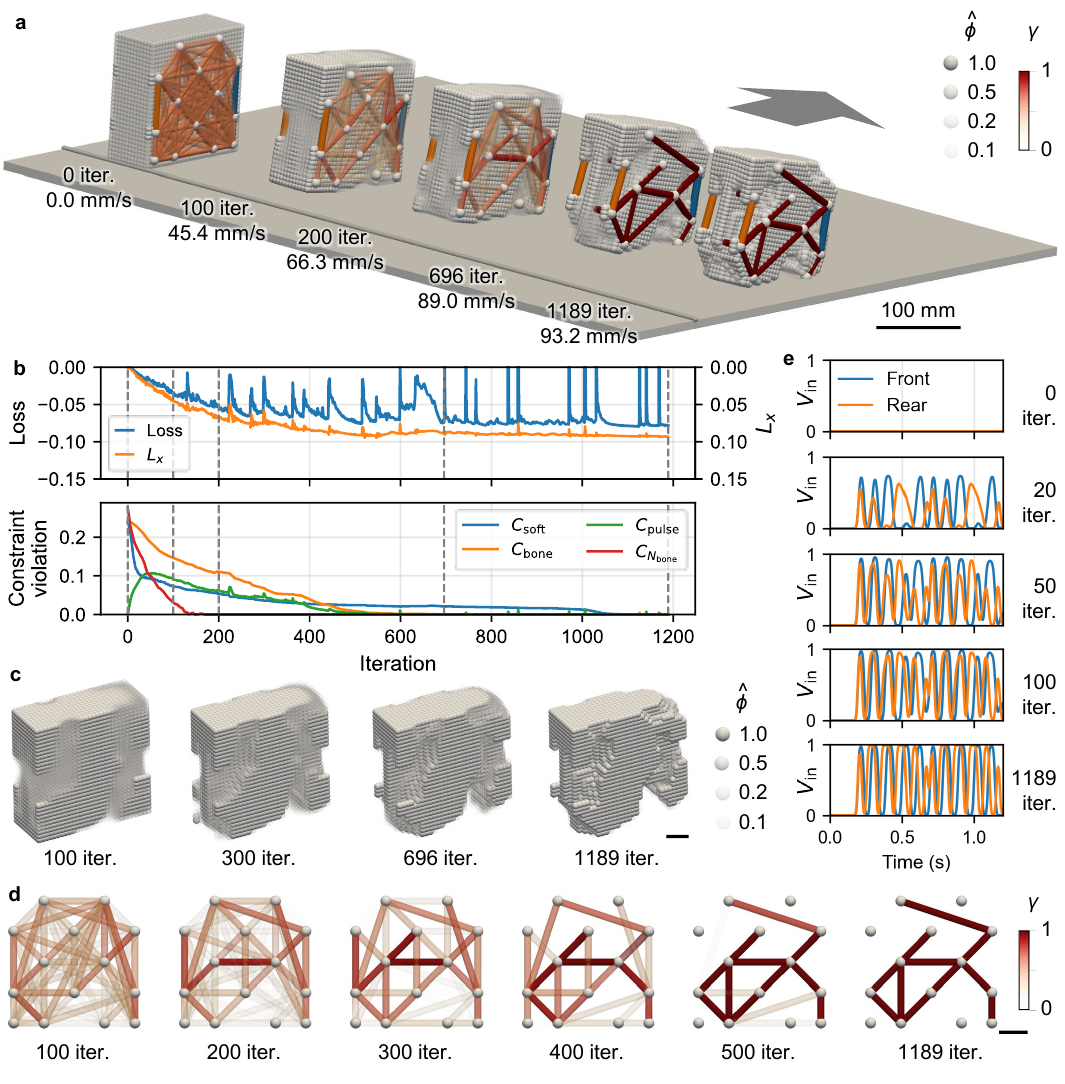}
\caption{Optimization history of the proposed framework.
\textbf{a} Representative intermediate designs from initialization to the final solution, showing the evolution of soft body, skeleton, and locomotion behavior.
\textbf{b} Optimization history of the loss, locomotion distance, and constraint terms, showing that locomotion performance improves while all constraints are progressively satisfied.
\textbf{c--e} Evolution of individual design components: \textbf{c} soft body morphology, \textbf{d} skeletal structure, and \textbf{e} actuation signal. Representative iterations are selected separately for each component to highlight their characteristic changes during optimization. The actuation is optimized at an early stage, followed by the development of the skeletal structure, while the soft body evolves more gradually throughout the process. Scale bars: 30 mm.}
    \label{fig2}
\end{figure}

We perform the optimization defined in the previous section and examine the resulting design process and behaviors (Fig.~\ref{fig2}, Supplementary Movie 1).
We now specify the physical and numerical settings used in the optimization.

The soft body is defined within a rectangular domain of size $0.15 \times 0.15 \times 0.06$~m.
The soft material is modeled with a Young's modulus of $E=0.144$~MPa and a Poisson ratio of $\nu=0.4$, corresponding to Lam\'e parameters $\mu=0.051$~MPa and $\lambda=0.206$~MPa. The soft material density $\rho$ is set to $1.07\times10^3$~kg/m$^3$. 
For the skeletal design, the nodes of a densely connected ground structure are placed on both sides of the soft body.
The skeletal elements are assumed to be linearly elastic with a Young's modulus of $3.0$~GPa and a rectangular cross section of $5.2 \times 1.0$~mm$^2$. The density of the skeletal material is set to $1.25\times10^3$~kg/m$^3$. 
This corresponds to the axial stiffness parameter
$
\kappa_s = 1.56\times10^{4}\ \mathrm{N/m}
$ per 1 m. 
Both the soft body and the skeletal layout are constrained to be symmetric with respect to the sagittal plane.

The front and rear solenoid actuators are modeled identically, using two endpoint masses of $16$~g and $75$~g, corresponding to the movable core (upper node) and the coil assembly (lower node), respectively, with a maximum contraction force of $F_{\max}=10$~N. 
The rest length of the actuator bar is $L_0 = 65$~mm, with a maximum stroke of $\Delta L = 15$~mm and an effective core length of $L_{\mathrm{core}} = 30$~mm. 
The axial actuator bars are assigned a low stiffness parameter of $\kappa_{\mathrm{free}} = 3.0 \times 10^{-1}$~N/m per 1~m under free-sliding conditions, which increases to $\kappa_{\mathrm{act}} = 3.0 \times 10^{8}$~N/m per 1~m when fully contracted to represent internal contact. 
Bridge bars are introduced to connect the skeletal structures across the actuator units. 
These elements are assumed to have a thicker rectangular cross section of $8.0 \times 8.0$~mm$^2$.

The simulation begins with a no-actuation phase from $t=0$ to $t_{\mathrm{start}}=0.2$~s, allowing the system to settle into a stable configuration under gravity.  
This is followed by an actuation phase from $t_{\mathrm{start}}=0.2$~s to $t_{\mathrm{end}}=1.2$~s, where a $0.5$~s actuation sequence is repeated twice.
The actuation over each $0.5$~s interval is discretized into a pulse sequence $w_{a,k}$ with a time step of $0.002$~s. 
The actuation sequence is repeated twice to mitigate transient inconsistencies between the beginning and end of a single cycle, thereby promoting more stable and repeatable locomotion behavior. To prevent tipping over in the narrow width direction, a slight tilt is introduced by modifying the gravity vector, corresponding to a $1^\circ$ inclination in the lateral direction.

All numerical simulations and optimizations were performed on a computer equipped with an NVIDIA A100 80~GB GPU, two AMD EPYC 7452 32-core processors. The program was implemented in Python 3.10.12 using Taichi 1.6.0~\cite{hu2019taichi}. The initialization and optimization loops were executed on the CPUs, whereas the MPM simulation and automatic differentiation were accelerated on the GPU using Taichi. The average computation time per iteration was approximately 178~s. Further details of the simulation and optimization settings are provided in ``Simulation and optimization details'' in Method section.

Figure~\ref{fig2}a shows representative intermediate designs illustrating the overall optimization process.
At the initial iteration, both the soft body and skeletal elements are initialized with intermediate densities, while the actuation signals are close to zero, resulting in no locomotion.
As the optimization proceeds, the soft body gradually adapts its morphology, while largely maintaining a dense structure with relatively limited void regions.
In contrast, the skeletal structure starts from a densely connected configuration and progressively eliminates unnecessary elements, leading to a sparse truss-like structure. During this process, the locomotion performance improves gradually.

By iteration 696, the skeletal layout has already converged to a configuration nearly identical to the final design, corresponding to the stage where the skeletal constraint $C_{\mathrm{bone}}$ is first satisfied.
Subsequent iterations up to the final solution (iteration 1189) primarily adjust the design to satisfy the constraints, with a further slight improvement in performance.
Specifically, the locomotion speed increases from 89.0~mm/s to 93.2~mm/s during this stage.

Figures~\ref{fig2}c--e show the evolution of the individual design components.
As shown in Fig.~\ref{fig2}c, the soft-body morphology evolves in a relatively consistent manner throughout the optimization.
Characteristic features, such as the formation of a small front limb, a larger posterior support region, and a gap between the actuator units, appear early and are progressively refined without major changes in overall topology.
By iteration 696, although the constraint $C_{\mathrm{soft}}$ is not yet enforced, relatively clear density distribution has already emerged.
In the final design, the projection further suppresses intermediate densities, leading to a more crisp density field.

In contrast, the skeletal structure in Fig.~\ref{fig2}d follows a more complex optimization trajectory.
Starting from a densely connected configuration, the elements gradually separate into high- and low-density groups.
However, regions with high density in early iterations do not necessarily persist in the final design, indicating a non-monotonic evolution.
Between iterations 400 and 500, the constraint progressively eliminates redundant elements, and by iteration 500 the skeletal structure is already close to the final configuration.

The actuation signal in Fig.~\ref{fig2}e converges much earlier than the other components.
In the early stage (up to around 20 iterations), a clear periodic pattern already emerges.
By around 100 iterations, the overall waveform structure is largely established, while the final optimized solution mainly refines its detailed shape.
This observation suggests that the optimization first identifies a suitable actuation frequency in the early stage, and subsequently adapts the soft body and skeletal structure to match it.
The actuation is represented as a sequence of Gaussian pulses with finite temporal width.
This parameterization imposes an inherent time scale on each contraction event, which limits the effective repetition frequency of actuation.
As a result, the optimization tends to favor actuation patterns near the upper end of the feasible frequency range, corresponding to rapidly repeated contraction cycles.
Further refinements in the waveform, such as the phase difference between front and rear actuators, continue until around 100 iterations, with the rear actuator generally leading the front.

\subsection*{Locomotion mechanism and experimental validation}
\begin{figure}[!tbp]
    \centering
    \includegraphics[width=0.72\linewidth]{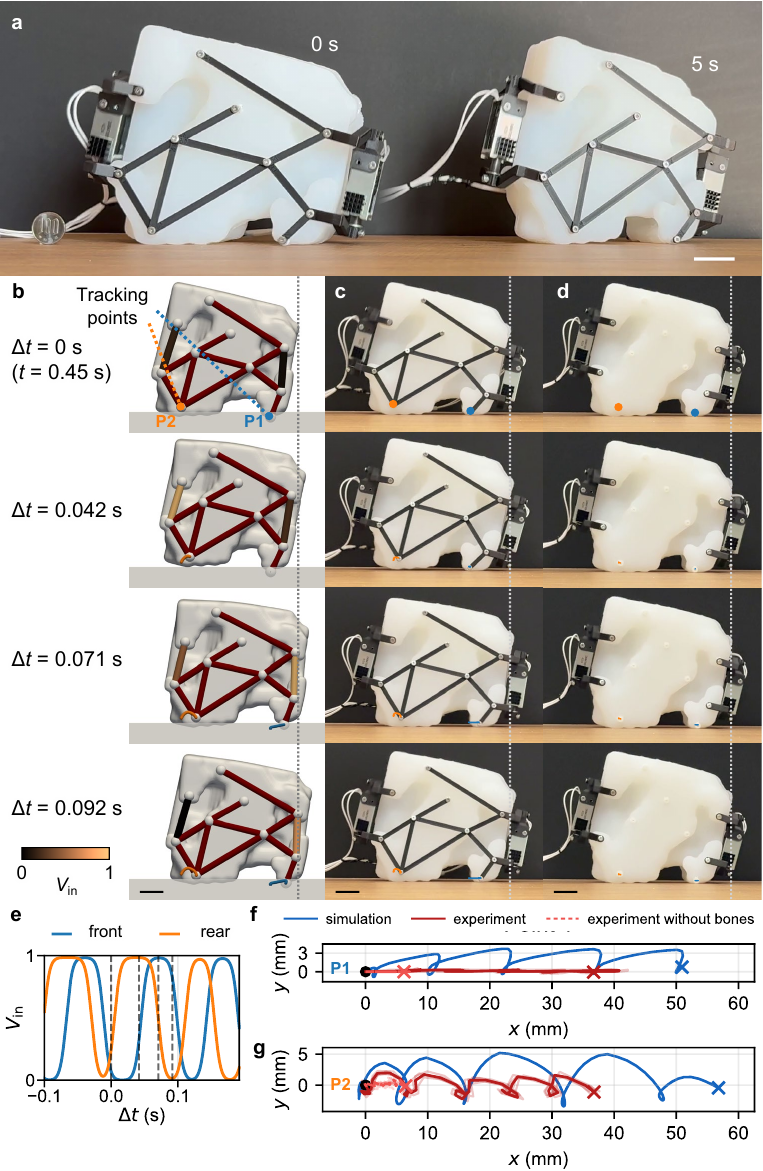}
    \caption{
Comparison between simulation and experiment, highlighting the role of the skeletal structure in locomotion.
\textbf{a} Snapshots of the fabricated robot during walking over 5~s, demonstrating sustained locomotion; a 100-yen coin is shown for reference.
\textbf{b--d} Representative snapshots over one actuation cycle for (\textbf{b}) simulation, (\textbf{c}) experiment, and (\textbf{d}) experiment without the skeletal structure, together with the trajectories of the front and rear nodes (P1 and P2).
In (\textbf{b}), the actuation state is visualized by the color of the actuator elements, with a corresponding color map.
The snapshots are taken at selected times within the cycle, and a 30~mm scale bar is shown in each image.
Both simulation and experiment exhibit similar motion patterns, including lifting of the rear leg, whereas the robot without the skeleton shows reduced vertical motion and less effective locomotion.
\textbf{e} Actuation signals corresponding to the cycle, with dashed lines indicating the snapshot timings.
\textbf{f,g} Trajectories of (\textbf{f}) the front node P1 and (\textbf{g}) the rear node P2 over 0.5~s.
Simulation, experiment, and experiment without skeleton are overlaid. Experimental results ($n=5$) are shown as thin lines with their mean in bold. The absence of the skeleton results in diminished vertical displacement and reduced forward progression. Scale bars: 30 mm.}
    \label{fig3}
\end{figure}

We analyze the locomotion mechanism of the optimized design and validate it through comparison between simulation and experiment (Fig.~\ref{fig3}, Supplementary Movie~2).
The fabrication generally follows Fig.~\ref{fig1}c, with a minor adjustment of the skeletal node positions.
The optimized design yields relatively short soft-body limbs, such that the front skeletal nodes extend beyond the soft body in the initial configuration.
In the simulation, these nodes become attached to the soft body after the initial settling phase under gravity.
Accordingly, the attachment points are adjusted based on the simulated configuration.

Figure~\ref{fig3}a shows the fabricated robot, which demonstrates sustained locomotion over 5~s. 
The robot travels 294~mm within this duration, despite its total mass of 1330~g.
This result indicates that the skeletal structure may play a role in effective load transmission and locomotion at this large scale.

Figures~\ref{fig3}b and \ref{fig3}c show representative snapshots of one actuation cycle in the simulation and experiment, respectively, together with the trajectories of two representative nodes, P1 and P2, corresponding to the front and rear legs.
The simulation consists of an initial no-actuation phase followed by a locomotion phase, and the snapshots are taken at selected time offsets from a reference time within this cycle.
The corresponding actuation signals are shown in Fig.~\ref{fig3}e, where the dashed vertical lines indicate the snapshot timings.
In the simulation (Fig.~\ref{fig3}b), locomotion is generated through an alternating actuation of the rear and front actuators.
Starting from a state where both legs are in contact with the ground, contraction of the rear actuator lifts the rear leg and moves it forward, resulting in an arched trajectory of the rear node.
As the actuation gradually shifts from the rear to the front ($\Delta t = 0.071$~s), the load distribution transitions from the front to the rear side.
During this phase, contraction of the front actuator lifts the front leg while advancing it forward, with the rear leg acting as a support point.
Subsequently, both actuators relax, returning the robot to a configuration similar to the initial state, and repetition of this cycle produces sustained locomotion.

The experimental results (Fig.~\ref{fig3}c) exhibit a similar sequence of motions, including lifting of the rear leg and forward progression.
For clarity, we primarily analyze the trajectories of the representative nodes.
The trajectory of the rear node P2 shows a similar arched pattern in both simulation and experiment, indicating consistent rear-leg motion.
In contrast, the trajectory of the front node P1 differs between the two cases: in the simulation, the node exhibits a slight lifting motion, whereas in the experiment it remains close to the ground.
This difference can be attributed to the contact modeling in the simulation.
In the simulation, contact is handled only for the soft body, and no contact constraint is imposed on the skeletal nodes, allowing them to penetrate the ground.
In the experiment, however, the nodes remain in contact with the ground while the surrounding soft material deforms, resulting in a flatter trajectory.
Despite these differences, both simulation and experiment consistently produce sustained locomotion.

Figure~\ref{fig3}d shows the experimental results without the skeletal structure.
Although the actuation of the solenoids is present, the resulting deformation of the soft body is limited and does not lead to clear locomotion.
The trajectories of the foot nodes also exhibit minimal displacement compared to the case with the skeleton.
This comparison highlights the role of the skeletal structure.
The externally attached skeleton enables load transmission between spatially separated regions of the body, allowing the actuation forces to be converted into coordinated, global deformation.
Without the skeleton, the actuation primarily induces local deformation of the soft body, which is insufficient to generate effective propulsion.

Figures~\ref{fig3}f and \ref{fig3}g show the trajectories of the front and rear nodes over a longer time window (0.5~s), corresponding to multiple actuation cycles.
Figure~\ref{fig3}f shows the trajectory of the front node $P_1$.
In the simulation, the node exhibits a sawtooth-like pattern, characterized by forward motion accompanied by gradual lifting, followed by a drop upon contact.
In contrast, the experimental trajectory shows little vertical motion, remaining close to the ground.
As discussed earlier, this difference is attributed to contact effects.
The experimental results without the skeleton show substantially reduced displacement, indicating that the skeletal structure plays an important role in achieving effective forward motion.
Figure~\ref{fig3}g shows the trajectory of the rear node $P_2$.
In the simulation, the trajectory follows an arched pattern, with the step length and vertical displacement varying across cycles, indicating a gradual change in motion amplitude.
In the experiment, the trajectory also exhibits an arched shape, but with more consistent step length and vertical motion across cycles.
The results without the skeleton show minimal displacement and negligible vertical motion, further supporting the role of the skeletal structure in generating effective locomotion.
The average forward speed in the experiments further highlights this difference.
With the skeleton, the mean speed was 69.669~mm/s (s.d. 1.118, $n=5$), whereas without the skeleton it was 19.906~mm/s (s.d. 1.178, $n=5$).

\subsection*{Effect of actuation frequency on locomotion}

Locomotion in compliant systems is often strongly influenced by frequency-dependent deformation and resonance-like behaviors, where specific actuation frequencies can enhance coordinated motion.
In soft--rigid hybrid systems, such behaviors emerge from the coupled dynamics between compliant materials and embedded skeletal structures, rather than from either component alone.
Motivated by this, we analyze how actuation frequency affects the deformation patterns and locomotion-relevant kinematics of the optimized design, and compare cases with and without the skeleton.

Actuation-induced deformation was analyzed using a finite element model implemented in COMSOL~6.3.
The soft body was modeled using the same material parameters as in the MPM simulations, with damping introduced via modal damping (loss factor $\tan\delta = 0.1$).
The skeletal structure was represented using beam elements with identical stiffness and mass parameters as in the XPBD model, assuming a uniform mass distribution.
A harmonic actuation force with a fixed magnitude of 1~N was applied to the actuator elements.
To evaluate the dynamic response, we performed a frequency sweep from 5 to 19~Hz with a resolution of 0.1~Hz.

To reproduce the coordinated actuation pattern observed in the optimized design, the phase relationship between the front and rear actuators was first identified from the optimized actuation signals and subsequently incorporated into the FEM analysis.
Specifically, the actuation signals obtained from the optimized design (Fig.~\ref{fig2}e) were analyzed in the frequency domain using Fourier analysis.
The front and rear actuator signals both exhibited a dominant frequency component at 10.00~Hz.
Based on this shared dominant frequency, the phase difference between the two signals was evaluated, revealing that the front actuator was activated after the rear actuator with a phase delay of $1.747$~rad.
This phase delay was then prescribed as the phase difference between the actuator inputs in the subsequent FEM vibration analysis.

\begin{figure}[!tbp]
    \centering
    \includegraphics[width=1.0\linewidth]{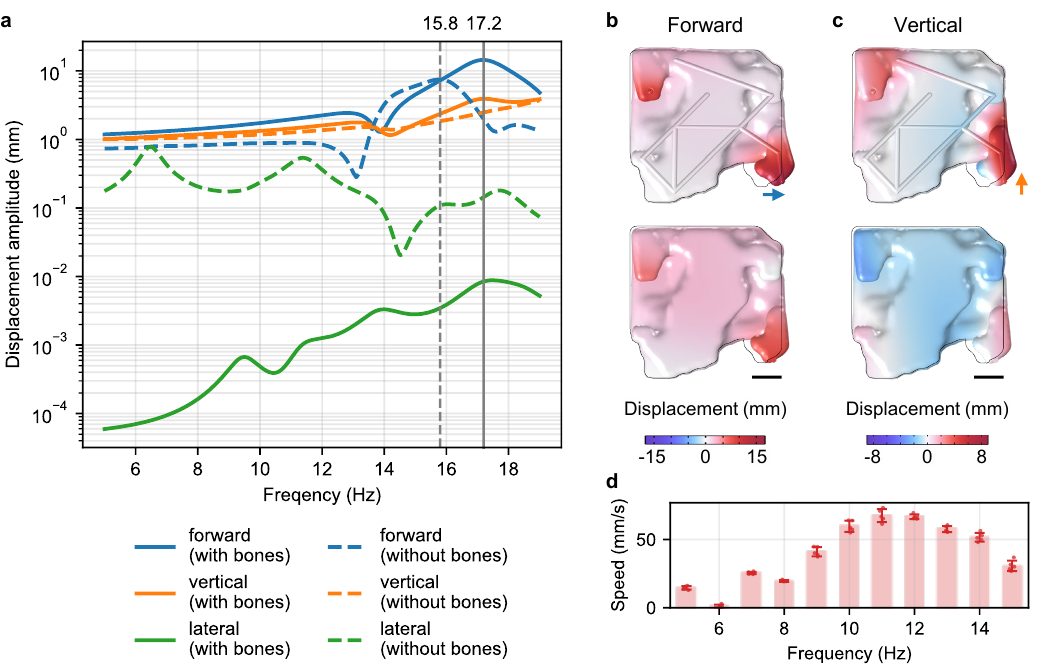}
    \caption{Frequency-dependent stride and locomotion performance.
\textbf{a}, Frequency response of the relative displacement between the front and rear limbs for forward, vertical, and lateral directions, with and without the skeleton.
The forward response shows peaks at 17.2~Hz (with skeleton) and 15.8~Hz (without skeleton).
The lateral response is reduced with the skeleton.
\textbf{b,c}, Displacement maps relative to the rear limb at the peak frequency for each case, shown for the forward (\textbf{b}) and vertical (\textbf{c}) directions, with and without the skeleton.
With the skeleton, the forward displacement shows a similar spatial pattern to the case without the skeleton, but with a larger amplitude.
In contrast, the vertical displacement exhibits qualitatively different behavior: with the skeleton, the front limb is lifted while the body remains relatively stable, whereas without the skeleton, deformation is more localized and the overall body tends to move downward. Scale bars: 30 mm.
\textbf{d}, Experimental locomotion speed as a function of actuation frequency ($n=5$, mean $\pm$ s.d.).
Peak performance occurs around 10--12~Hz, while locomotion is strongly degraded at 6~Hz.}
    \label{fig4}
\end{figure}

The response was quantified using the relative displacement between the front and rear limbs, computed as the difference in their averaged positions over predefined regions (for details, see Supplementary Fig.~1).
The frequency response shows that the presence of the skeletal structure leads to larger deformation amplitudes over a wide range of frequencies, with peaks in the forward direction at 17.2~Hz (with the skeleton) and 15.8~Hz (without the skeleton) (Fig.~\ref{fig4}a).
For both the forward and vertical directions, the peak amplitudes are generally higher in the presence of the skeleton.
In addition, the lateral displacement is reduced in the presence of the skeleton, indicating suppression of out-of-plane deformation.

To further examine the deformation patterns, spatial distributions of displacement at the peak frequency are visualized in Fig.~\ref{fig4}b,c (for details, see Supplementary Movie~3).
For each case, the phase is selected such that the relative displacement between the front and rear limbs is maximized, and the corresponding real displacement field is shown.
In the forward direction (Fig.~\ref{fig4}b), the overall deformation pattern is similar between the two cases, but the skeletal structure amplifies the displacement.
In contrast, the vertical displacement (Fig.~\ref{fig4}c) shows qualitatively different behavior, with pronounced lifting of the front limb only in the skeletal case.

This difference can be attributed to the structural role of the skeleton.
In the region around the rear limb, the skeleton forms a truss-like configuration that reinforces the soft body, providing a more stable support.
In the front region, the skeletal elements form a triangular linkage that includes the actuator, where the contraction of the actuator drives the lower joint to move both forward and upward.
This mechanism converts the actuator stroke into coordinated lifting and propulsion of the front limb, which is not achieved without the skeleton.

To examine how these frequency-dependent responses relate to actual locomotion, we conducted experiments in which the same on--off actuation pattern (repeated four times over 2~s in the original setting) was compressed or stretched in time to vary the effective actuation frequency.
The locomotion speed was evaluated as the traveled distance normalized by the actuation duration.
Each condition was tested over $n=5$ trials.
Representative videos at each frequency are provided in Supplementary Movie~3.
The experimental results (Fig.~\ref{fig4}d) show a clear dependence of locomotion performance on actuation frequency, with a peak around 10--12~Hz, where the highest speed is observed at 11~Hz, followed by 12~Hz and 10~Hz, which corresponds to the frequency obtained through the optimization.
As the frequency deviates from this range, the locomotion performance gradually decreases.
Notably, at 6~Hz, the robot exhibits almost no forward motion, resulting in a pronounced drop in performance, whereas neighboring frequencies such as 5~Hz and 7~Hz still produce measurable locomotion.

Direct quantitative comparison between the vibration analysis and the experimental locomotion results is not straightforward, because the analyzed relative leg motions are indirectly related to locomotion performance and the FEM simulation does not include a part of phenomena such as gravity or ground contact.
Nevertheless, comparing these results with the frequency response (Fig.~\ref{fig4}a), we observe qualitatively similar trends, characterized by a single peak followed by a gradual decrease, as well as a distinct low-response region at lower frequencies. 

Taken together, these observations show that locomotion performance depends strongly on actuation frequency, with specific frequency ranges leading to effective locomotion.
Notably, the frequency identified by the optimization (10~Hz) lies close to the experimentally observed performance peak.
This suggests that the optimization process selects actuation frequencies that align with favorable dynamic responses of the system.
As a result, the morphology of the soft body, the skeletal layout, and the actuation timing are jointly tuned to exploit these underlying dynamics.

\section*{Discussion}

This study proposes a framework for automatically designing soft--rigid hybrid robots and highlights the role of combining soft and rigid components in locomotion.
By simultaneously optimizing the soft-body morphology, skeletal structure, and actuation, the method enables the generation of functional soft--rigid robots without manual design, and its validity is confirmed through experimental realization.

A central contribution of this work lies in the treatment of rigid structural elements within the design framework.
The design results reveal that the soft body and the skeleton play fundamentally different roles: the soft body primarily contributes to deformation modes and thus benefits from maintaining high geometric flexibility, whereas the skeletal structure governs connectivity and load transmission across the body.
This observation motivates the use of different representations for the two components, namely a continuum-based formulation for the soft body and a ground-structure representation for the skeleton.

Based on this perspective, we introduce a formulation that enables gradient-based optimization of stiff structural elements by incorporating an XPBD-based simulator.
Rather than treating rigid components as ideal rigid bodies, the proposed approach represents them using constraint-based formulations compatible with differentiable simulation, allowing efficient optimization without solving large systems.
This makes it possible to handle a large number of design variables associated with skeletal connectivity at a relatively low computational cost.
The rigid formulation is coupled with a differentiable Material Point Method (MPM) for the soft body, resulting in a unified framework that captures both large deformation and structural connectivity within a single optimization process.

The optimized structures exhibited both biologically reminiscent and engineering-driven characteristics.
In the front limb region, the optimized connectivity formed knee-like structures, whereas the rear body developed triangular truss-like reinforcement.
Despite these structural differences from biological organisms, the optimized robot exhibited alternating front–rear motion reminiscent of animal locomotion patterns.
This suggests a potential advantage of optimization-based design over direct biological imitation, as it can adapt structures to engineering requirements while retaining functional behaviors.

From a design perspective, the proposed framework also extends existing approaches for automated soft-robot design.
While many prior methods focus on robots composed solely of soft materials, the present formulation allows the incorporation of rigid components within the design space.
This capability is not limited to skeletal structures for locomotion, but can also accommodate practical design requirements, such as embedding actuators, rigid supports, or functional components.
As a result, the framework can be interpreted not only as a tool for designing soft--rigid locomotion systems, but also as a more general design approach for soft robotic systems with heterogeneous material composition.

From a practical perspective, the fabricated robot operates at a relatively large scale (approximately 150~mm in size and 1330~g in mass), while still exhibiting effective soft-body deformation contributing to locomotion.
While soft robotic systems are often realized at relatively small scales, the present results suggest that incorporating rigid load-bearing elements may enable scalable designs, consistent with biological strategies for supporting increased body size.

Several directions can be explored to extend the present framework.
First, the design space can be further expanded.
In the current formulation, the nodal positions of the ground structure are fixed, and optimizing their placement would provide additional geometric flexibility.
Moreover, the skeletal elements are modeled as pin-jointed members, and extending the formulation to represent moment-transmitting structures could enable richer mechanical behaviors.
While some of these effects can be approximated through triangulated connectivity, an explicit treatment of moment transfer may further enhance the design space.
Second, the placement of actuators is expected to play a substantial role in determining locomotion performance.
In this work, actuator locations are predefined for simplicity and manufacturability, but extending the framework to optimize actuator placement could lead to more efficient designs.
Such an extension may be achieved by integrating actuator variables into the ground-structure formulation.
Finally, the current approach relies on gradient-based optimization and is therefore inherently local.
Incorporating complementary strategies, such as evolutionary algorithms or machine learning-based approaches, may enable broader exploration of the design space and facilitate the discovery of more diverse solutions.

\section*{Methods}
\subsection*{Soft body modeling}
The soft domain is governed by conservation laws of mass and momentum,
\begin{equation}
\frac{D\rho}{Dt} + \rho \nabla\cdot \boldsymbol{u} = 0,
\qquad
\rho\frac{D\boldsymbol{u}}{Dt} = \nabla\cdot\boldsymbol{\sigma} + \rho\boldsymbol{g},
\end{equation}
where $\rho$ is the density of soft body, $t$ is the time, $\boldsymbol{u}$ is the velocity, $\boldsymbol{\sigma}$ is the Cauchy stress, and $\boldsymbol{g}$ is gravity.

We used a compressible neo-Hookean constitutive law with an additional viscous term,
\begin{equation}
\boldsymbol{\sigma} =
\frac{1}{J}\left[
\mu(\hat{\phi})\,\boldsymbol{F}\boldsymbol{F}^{\mathsf{T}}
+\big(\lambda(\hat{\phi})\ln J-\mu(\hat{\phi})\big)\boldsymbol{I}
\right]
+
\eta\left(
\nabla\boldsymbol{u}+(\nabla\boldsymbol{u})^{\mathsf{T}}
-\frac{2}{3}(\nabla\cdot\boldsymbol{u})\boldsymbol{I}
\right),
\end{equation}
where $\boldsymbol{F}$ is the deformation gradient, $J=\det\boldsymbol{F}$, and $\boldsymbol{I}$ is the identity matrix. 
The parameters $\mu(\hat{\phi})$ and $\lambda(\hat{\phi})$ are the Lam\'e parameters, and $\eta$ is a fictitious viscosity coefficient. 
The first term represents hyperelastic stress, while the second term is a fictitious viscous stress used to capture dissipative effects in a simplified manner. 

\subsection*{XPBD formulation}
The embedded skeleton and actuator structures were modeled as distance constraints between connected nodes within the XPBD framework.
For each bar $s$ connecting nodes $a(s)$ and $b(s)$, we define the distance-constraint residual as
\begin{equation}
g_s(\boldsymbol{x}) := \|\boldsymbol{x}_{a(s)}-\boldsymbol{x}_{b(s)}\| - L_s^{\mathrm{ref}}.
\label{eqn_constraint}
\end{equation}
where $L_s^{\mathrm{ref}}$ denotes the rest length of the bar used in the constraint. 
Given predicted positions obtained from explicit time integration of inertial and external forces, we perform XPBD projection as
\begin{align}
\Delta\Lambda_s &=
\frac{-g_s(\boldsymbol{x}) - \tilde{\alpha}_s\Lambda_s}{\frac{1}{m_{a(s)}}+\frac{1}{m_{b(s)}}+\tilde{\alpha}_s},
\\
\tilde{\alpha}_s &= \frac{1}{\Delta t^2}\frac{L_s^{\mathrm{ref}}}{\kappa_s}, \\
\boldsymbol{x}_{a(s)} &\leftarrow \boldsymbol{x}_{a(s)} + \frac{1}{m_{a(s)}}\Delta\Lambda_s \boldsymbol{n}_s,
\\
\boldsymbol{x}_{b(s)} &\leftarrow \boldsymbol{x}_{b(s)} - \frac{1}{m_{b(s)}}\Delta\Lambda_s \boldsymbol{n}_s,
\end{align}
where $m_i$ is the node mass, $\Delta t$ is the time step, $\Lambda_s$ is the Lagrange multiplier, and $\boldsymbol{n}_s$ is the unit direction vector connecting two nodes. 

In standard XPBD, the constraint is solved iteratively within each time step.
However, we instead adopt a substepping strategy with a single projection per step, which achieves faster constraint convergence at a similar computational cost without sacrificing solution consistency~\cite{macklin2019small}.

\subsection*{Buckling-aware stiffness model}

To account for potential buckling of slender bars, we incorporate a simple buckling-aware stiffness model based on Euler's critical load.

The critical load for each bar is estimated as
\begin{equation}
P_{\mathrm{cr},s} = \frac{\pi^2 E_s I_s}{(K_b L_s^{\mathrm{ref}})^2},
\end{equation}
where $E_s$ is the effective Young's modulus, $I_s$ is the second moment of area, and $K_b$ is the effective length factor.

The compressive load is approximated as
\begin{equation}
P_{\mathrm{est},s} = \frac{\kappa_s}{L_s^{\mathrm{ref}}}(L_s^{\mathrm{ref}}-\ell_s),
\end{equation}
where $\ell_s=\|\boldsymbol{x}_{a(s)}-\boldsymbol{x}_{b(s)}\|$ is the current bar length.

When the bar is in compression and the estimated compressive load exceeds the critical load, 
the effective stiffness is reduced by limiting the compressive response as
\begin{equation}
\kappa_s^{\mathrm{eff}}=
\begin{cases}
\max\!\left(
\kappa_{\min,s},\,
\min\!\left(
\kappa_s,\,
\dfrac{P_{\mathrm{cr},s}L_s^{\mathrm{ref}}}{L_s^{\mathrm{ref}}-\ell_s}
\right)\right),
& \text{if } \ell_s<L_s^{\mathrm{ref}}\ \text{and}\ P_{\mathrm{est},s}\ge P_{\mathrm{cr},s},\\[6pt]
\kappa_s, & \text{otherwise}.
\end{cases}
\end{equation}
where $\kappa_{\min,s}$ is a lower-bound stiffness introduced to avoid numerical degeneracy.

This model provides a simple approximation of buckling effects without explicitly resolving instability modes.

\subsection*{Solenoid actuator model}
To reproduce the stroke-dependent mechanical response of the physical solenoid mechanism, the axial actuator bars were modeled with state-dependent stiffness. 

During free sliding, the axial bars were assigned a low stiffness $\kappa_{\mathrm{free}}$ with the reference length fixed at the installed length $L_0$. 
Once the actuator length $\ell$ reached the contraction limit $L_0-\Delta L$, the stiffness was switched to $\kappa_{\mathrm{act}}$ and the reference length was changed to $L_0-\Delta L$, where $\Delta L$ denotes the actuator stroke.

The actuation force in Eq.~(\ref{eqn_stroke_force}) depends on the stroke-dependent attenuation factor $\eta_{\mathrm{stroke}}(\ell)$, defined as
\begin{equation}
\eta_{\mathrm{stroke}}(\ell)=
\begin{cases}
1,
& \ell \le L_0-\Delta L, \\
1-\dfrac{\ell-(L_0-\Delta L)}{L_{\mathrm{core}}},
& L_0-\Delta L < \ell < L_0-\Delta L+L_{\mathrm{core}}, \\
0,
& \ell \ge L_0-\Delta L+L_{\mathrm{core}}.
\end{cases}
\end{equation}
Here, $L_{\mathrm{core}}$ denotes the effective length of the iron core. 
The attenuation factor takes its maximum value of $1$ when the actuator is fully contracted, decreases linearly as the core exits the coil, and becomes zero once the core completely disengages from the coil. 
Further extension beyond this point does not generate additional force.

\subsection*{Simulation setup}
The simulation is based on a material point method (MPM), consisting of particle-to-grid (P2G) transfer, grid-based update, and grid-to-particle (G2P) transfer. Supplementary Fig.~2 shows the flowchart of one forward step in the simulation.
In the P2G step, particle mass and momentum are transferred to the background grid.
The grid state is then updated under external forces and boundary conditions, and the updated grid velocity is transferred back to particles in the G2P step.

On top of this MPM framework, the dynamics of the embedded bar network are advanced using an extended position-based dynamics (XPBD) formulation.
Starting from the current nodal positions and velocities, provisional positions are predicted, and constraint projection is applied to enforce bar-length constraints and boundary conditions, yielding updated nodal motion for the next step.
The coupling between MPM and XPBD is realized through momentum exchange on the grid.
Specifically, the constraint-corrected nodal motion is converted into an effective nodal velocity and transferred to the MPM grid as additional momentum contributions.
Each time step consists of this combined MPM update and XPBD-based correction.

For MPM simulation, the domain is embedded in a cubic background grid of size $0.8 \times 0.8 \times 0.8$~m, discretized into $80 \times 80 \times 80$ cells. 
This gives a grid spacing of $dx=0.01$~m. 
MPM particles are placed with a spacing of $dx/2=0.005$~m.

The MPM and XPBD solvers share the same time step of $dt = 2 \times 10^{-5}$~s. For XPBD, constraint projection is performed once per time step.
To reduce memory usage in the differentiable simulation, the forward trajectory is divided into checkpoint intervals of 250 steps, and intermediate states are recomputed during backpropagation following the checkpointing strategy~\cite{hu_difftaichi_2020}. 

\subsection*{Optimization setup}

The optimization procedure is illustrated in Supplementary Fig.~3.

For the soft-body optimization, density filtering and projection are applied. 
The filter radius is set to $0.02$~m, and the weighting profile follows a power-law function with exponent 3. 
The filtered field is then projected to the physical density field using a smooth projection function~\cite{yuhn20234d} with $\beta = 8$. 
For the actuation, each pulse is defined as a Gaussian function with an amplitude coefficient of $0.2$ and a standard deviation of $0.01$~s, and the actuation waveform is constructed as the superposition of these pulses.

For a given set of design variables, including the soft-body shape, the skeletal bar network, and the actuation pattern, the locomotion performance and constraint values are evaluated. Next, the gradients with respect to all variables are computed via backward simulation. 

To promote near-binary designs, quadratic constraints are imposed on the soft-body density, skeletal density, and actuation variables. 
The upper bounds are set to $0.0125$ for the soft-body density and $0.0025$ for the actuation variables, corresponding to tolerances of $5\%$ and $1\%$, respectively.
The skeletal density bound is set to $0.0125/180$, where the number of skeletal elements excluding actuators is 180 in the present design setting. With this setting, even a single skeletal element taking an intermediate value (e.g., $\gamma = 0.5$) violates the bound, thereby strongly suppressing intermediate skeletal elements.
In addition, the number of skeletal bars is constrained to be at most 40, which defines an upper bound on the average skeletal density of $40/180 \approx 0.217$.

To handle constraints, an augmented Lagrangian formulation is employed.
The optimization is initialized in a nearly unconstrained regime with small penalty parameters, which are increased adaptively when the objective function is stationary. Convergence is assessed based on the relative change in the objective value, where the optimization is considered stationary when the relative difference between the averages over two consecutive windows of five iterations falls below $10^{-3}$.
For the update of Lagrange multipliers, the binarization constraint for the soft body, $C_{\mathrm{soft}}$, is enforced only after the corresponding skeletal constraint, $C_{\mathrm{bone}}$, is sufficiently satisfied, in order to avoid premature discretization of the soft structure.

The design variables are updated using the Adam optimizer~\cite{kingma_adam_2015}, with learning rates of $0.02$ for the soft-body variables, $0.01$ for the skeletal variables, and $0.02$ for the actuation variables.

\subsection*{Fabrication of soft--rigid hybrid robot}

The soft body was fabricated using a platinum-cure silicone elastomer (Ecoflex 00-50, Smooth-On Inc.). A two-part mold consisting of left and right halves was designed and produced via 3D printing. The silicone was cast separately into each half of the mold and subsequently bonded by applying an additional layer of uncured silicone at the interface. The assembled parts were then cured in an oven at 70~$^\circ$C for 15~min to promote crosslinking. After curing, the mold was destructively removed to obtain the soft body. The fabricated soft body exhibited adhesive contact with the ground, leading to undesired sticking during locomotion. To mitigate this effect, a dot-pattern surface texture was introduced on the foot regions by selectively applying adhesive (Super X Gold, Cemedine Co., Ltd.). Details of this surface pattern are provided in Supplementary Fig.~4.

The skeletal components were fabricated by fused filament fabrication using polylactic acid (PLA; Bambu PLA Matte, Bambu Lab) on a desktop 3D printer (P2S, Bambu Lab). Each skeletal element had a rectangular cross-section of 5.2~mm $\times$ 1~mm, with lengths determined by the optimized design. Connection points were designed with through-holes of 3.2~mm diameter for mechanical fastening.

Actuation was provided by pull-type solenoids (CA08470090, Takaha Kiko Co., Ltd., 9~$\Omega$ model). The solenoids were driven using a multi-controller (Multi Controller A, Takaha Kiko Co., Ltd.) in combination with a microcontroller (Arduino Uno R4, Arduino), with a supply voltage of 20~V. The solenoid mounting brackets (bridge bars) were fabricated using a fiber-reinforced 3D printer (Onyx Pro, Markforged) with a nylon-based composite filament (Onyx, Markforged).

To enable assembly, nylon nuts (M3) were embedded into the soft body using an adhesive (Super X Gold, Cemedine Co., Ltd.), and the skeletal elements and actuator fixtures were subsequently attached via screw fastening.

\section*{Author contributions}
Conceptualization was carried out by H.K. and T.N.
Methodology was developed by H.K., Y.T., C.Y., Y.S., S.T., A.K., and T.N.
Theoretical and numerical investigations were carried out by H.K., Y.T., and S.T.
Experiments were performed by H.K.
The manuscript was drafted by H.K. and revised by Y.T., C.Y., Y.S., S.T., A.K., and T.N.
The research project was supervised by T.N.

\section*{Competing interests}
The authors declare no competing interests.





\bibliography{sn-bibliography}

\end{document}